\documentclass[a4paper]{mem}
\usepackage{natbib}
\usepackage{graphicx}
\usepackage[a4paper]{hyperref}
\usepackage{txfonts}
\begin{document}
   \title{Gamma-ray Bursts observed by XMM-Newton
}

   \author{P.T. O'Brien\inst{1}, 
   J.N. Reeves\inst{2},
           D. Watson\inst{3},
          J. Osborne\inst{1} \and R. Willingale\inst{1},
}

   \offprints{P.T. O'Brien, \\ \email{pto@star.le.ac.uk} }

   \institute{Department of Physics \& Astronomy, University of Leicester, 
University Road, Leicester, LE1 7RH, U.K. \\ 
              \and  Laboratory for High Energy Astrophysics, NASA
Goddard Space Flight Center, Code 662, Greenbelt, MD 20771,
 USA\\
              \and Astronomical Observatory, Neils Bohr Institute for
Astronomy, Physics and Geophysics, University of Copenhagen,
Juliane-Maries, Vej 30, DK-2100, Copenhagen, Denmark
             }

   \abstract{ Analysis of observations with XMM-Newton have made a
significant contribution to the study of Gamma-ray Burst (GRB) X-ray
afterglows. The effective area, bandpass and resolution of the EPIC
instrument permit the study of a wide variety of spectral features. In
particular, strong, time-dependent, soft X-ray emission lines have
been discovered in some bursts. The emission mechanism and energy
source for these lines pose major problems for the current generation
of GRB models. Other GRBs have intrinsic absorption, possibly related
to the environment around the progenitor, or possible iron emission
lines similar to those seen in GRBs observed with BeppoSAX. Further
XMM-Newton observations of GRBs discovered by the Swift satellite
should help unlock the origin of the GRB phenomenon over the next few
years.

   }
   \authorrunning{P.T. O'Brien et al.}
   \titlerunning{Gamma-ray bursts observed by XMM-Newton}
   \maketitle

\section{Introduction}

Gamma-ray Bursts (GRBs) are thought to signal the birth of a black
hole which is for a short period (few to few tens of seconds) fed by a
surrounding disk or torus. The progenitor is probably a massive
collapsing star, possibly in a binary system, which, as the system
collapses, release enormous amounts of energy ($\sim 10^{52}$ erg)
which powers a relativistic jet. The jet tunnels its way out
of the progenitor star envelope and then interacts with any surrounding
material. This interaction results in an ``afterglow'' which emits
strongly from X-ray to radio wavelengths before fading from view. GRBs
can be crudely divided into ``short'' and ``long'' bursts depending on the
length of the gamma-ray pulse, with the division occurring at about 2
seconds. Virtually all multi-wavelength data, including all that
discussed here, has been obtained for long bursts. A recent review of
the GRB phenomena is given in \cite{zhang}.

Following the discovery of X-ray afterglows, it was soon realised that
the X-ray emission provides a probe of the extreme conditions in the
GRB jet. The time-dependent X-ray spectrum acts as a monitor of the
density, temperature and emission mechanism, if data of sufficient
quality can be obtained. BeppoSAX observations of GRB afterglows
revealed emission features usually attributed to iron K$\alpha$ (e.g.
\citealt{antonelli}). The large effective area of the XMM-Newton
EPIC instrument affords the opportunity to search for weaker lines,
study their time-dependence and to explore the rich, soft X-ray band.
Over the last few years XMM-Newton has observed a number of GRBs most
of which are described below. Given the operational constraints,
XMM-Newton observations have not begun earlier than $\sim 0.2$ days
after the GRB (in the GRB rest-frame), but this is still fast enough
for EPIC X-ray spectroscopy of most bursts.

\section{GRB observations}

The list of GRBs we have analysed is shown in Table 1. GRBs are named
in terms of the year+month+day they were detected. A letter is added
if several candidates occurred on the same day. The claimed spectral
features are listed and discussed below along with the derived ``X-ray
redshift''. Only two of the GRBs studied have a known optical redshift
to date.

\begin{table*}
 \centering
 \begin{minipage}{140mm}
\caption{Summary of the XMM-Newton GRB observations discussed in
this paper. $^a$Rest-frame time interval (in days) over which lines
seen. Observing epoch given in brackets if no lines detected.
$^b$X-ray redshift. $^c$Optical redshift. $^d$ Significance level for
line detection. $^e$ Publication reference. }

\begin{tabular}{@{}lcccccl@{}}
GRB & Rest-frame interval$^a$ & X-ray Lines & $z_x^b$ & $z_{opt}^c$ &
Sig.$^d$ & Ref.$^e$ \\  
011211 & $<0.15 - 0.17$ & H-like Si, S, Ar & $1.86 \pm 0.07$ & 2.140 &
99.97\% & 1, 2 \\
001025A & $<0.12 - >1.4$ & (H-like Mg, Si, S) & $0.53\pm0.03$ & &
99.87\% & 3 \\
010220 & $<0.31 - >0.56$ & ``Fe'' & $1.0\pm0.05$ & & 99.84\% & 3 \\
020322 & (0.22 - 0.34) & absorption & 1.8(?) & & & 4 \\
030227 & $ 0.18 - >0.23$ & H-like Si, S, He-like Ca & $1.39\pm 0.05$
& & $99.85\%$ & 5\\
030329 & (36.3 and 51.7) & none & & 0.1685 & & 6, 7 \\ \ \\
\end{tabular}
{References: 1. \cite{reeves2}; 
2 \cite{reeves3}; 
3. \cite{watson2a}; 
4. \cite{watson2b}; 
5. \cite{watson3}; 
6. \cite{willingale}; 
7. \cite{tiengo}
}
\end{minipage}
\end{table*}

   \begin{figure}
   \centering
   \includegraphics[angle=-90,width=6.5cm]{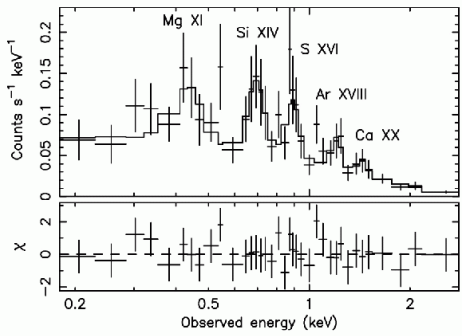}
      \caption{XMM-Newton EPIC PN spectrum of GRB011211 for the first 5 ksec
of the observation (\citealt{reeves2}). Top: 
count-rate spectrum with best-fit thermal
model and Galactic absorption. Bottom: chi-squared fit residuals. 
The identified emission lines are marked.
              }
\label{fig1}
   \end{figure}

   \begin{figure}
   \centering
   \includegraphics[angle=-90,width=6.5cm]{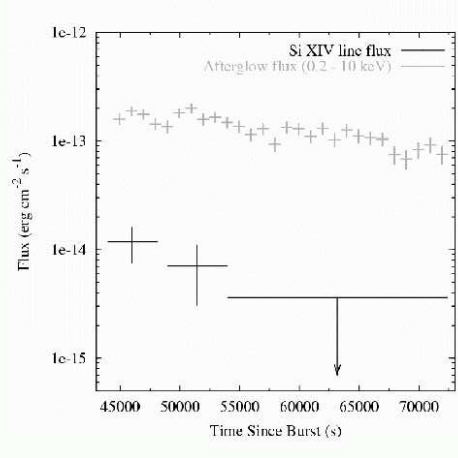}
      \caption{The decay in the flux of the continuum and the Si XIV
line during the XMM-Newton observation of GRB011211. The line is only
detected early in the observation.
              }
\label{fig2}
   \end{figure}

\subsection{GRB011211}

GRB011211 was first detected by BeppoSAX at 19:09:21 (UT) on 2001
December 11, and was the longest burst (270 seconds) observed by
BeppoSAX. XMM-Newton observations started at 06:16:56 on 2001 December
12, some 11 hours after the gamma-ray flash (\citealt{santos}).
The optical redshift of the burst was determined by \cite{holland}
to be $z = 2.140 \pm 0.001$, implying an isotropic luminosity
of $5\times10^{52}$ erg.

Our analysis of the XMM-Newton data is described in detail in Reeves et
al. (2002; 2003). The most important discovery was the serendipitous
observation of fading soft X-ray emission lines. These lines are most
clearly visible in the first 5 ksec of the observation (Figure 1) and
then fade faster than the continuum (Figure 2). The lines are from hydrogenic
states of Mg, Si, S, Ar and Ca. Fitted together using Gaussian
profiles to represent the lines, a weighted mean X-ray line redshift
can be derived of $z = 1.86\pm 0.07$. This is significantly different
from the optical redshift, and implies an outflow velocity of
approximately 0.1c.

The presence of decaying, soft X-ray lines was not predicted by GRB
models and illustrates the potential of XMM-Newton for such work.
Equally intriguing is the lack of iron emission in GRB011211. The
best-fit model for the emission is, surprisingly, an
optically-thin thermal plasma. Although unlikely to be fully
physically realistic, this model allows for an estimate of the
abundances and ejected mass. For the light elements an abundance some
10 x Solar is required but $<1.4$ x Solar for iron. The ejected mass is
4--20 Solar masses.

\subsection{GRB030227}

Following the discovery of lines in GRB011211 we searched for such
features in all GRBs observed by XMM-Newton. The best case, and indeed
the strongest of all, is for GRB030227. This GRB was first detected by
Integral with the XMM-Newton observation starting some 8 hours after
the burst. \cite{watson3} discuss the data and show that soft
X-ray emission lines are strongly detected near the end of the
observation (Figure 3). The lines are rising while the continuum is
decaying, although there is a small possible rise in the continuum
during in the last $\sim 10$ ksecs (Figure 4). The X-ray line spectrum
in GRB030227 is remarkably similar to that of GRB011211, showing
hydrogen and/or helium-like states of Mg, Si, S, Ar and Ca. Again
there is no evidence for Fe and neither Co or Ni are detected.

The best-fit redshift derived from the X-ray lines is
$z=1.39^{+0.03}_{-0.06}$. Unfortunately there is, as yet, no optically
determined redshift for this burst so we cannot search for a velocity
shift. Nevertheless, the similarity with GRB011211 gives us confidence
that a similar phenomena is occurring in both objects. Again, the soft
X-ray line spectrum in GRB030227 can be well fitted by a thermal
plasma, giving a minimum abundance of 24x Solar for the light elements
compared to $<1.6$ and $<18$ for Fe and Ni respectively. Watson et al.
also note that the continuum energy required to produce the 
X-ray line emission (assuming isotropic emission) is, at the very least,
within a factor of 2 of the gamma-ray energy. The time-delayed
emergence of bright line emission implies a continuing injection of
energy.

   \begin{figure} \centering
\includegraphics[angle=-90,width=6.5cm]{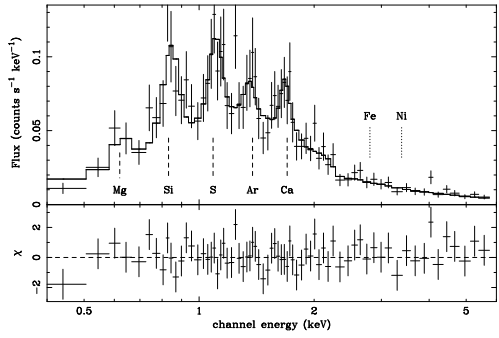} \caption{XMM-Newton
EPIC PN spectrum of GRB030227 for the last 11 ksec of the observation
(\citealt{watson3}).
Top: count-rate spectrum fitted with a powerlaw, Galactic absorption
and five Gaussian emission lines. Bottom: chi-squared fit residuals.
The identified emission lines are marked at their expected energies
for a redshift of $z=1.39$. } 
\label{fig3}
\end{figure}

   \begin{figure}
   \centering
   \includegraphics[angle=0,width=6.5cm]{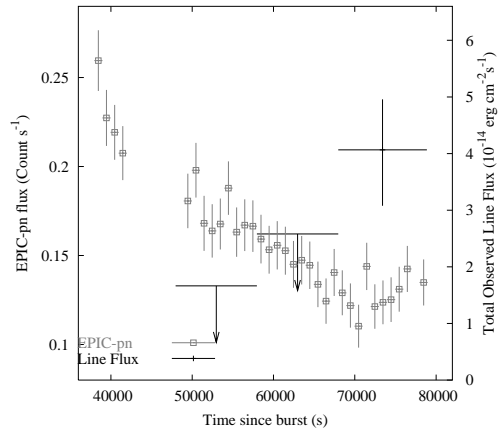}
      \caption{
The light-curve of GRB030227 during the XMM-Newton observation.
The continuum ((0.2--10) keV) decays steadily during the observations
whereas the emission lines become detectable only at the end.              }
\label{fig4}
   \end{figure}

The observed spectra of GRB011211 and GRB030227 suggest either little
production of heavy elements, in contrast to the usual predictions,
or that somehow the heavy elements are missing from the observed
ejecta (\citealt{watson3}). The expected abundance patterns depend
strongly on the physical conditions, geometry and dynamics of the
nucleosynthesis site which could be the progenitor, the disk/torus
surrounding the black hole and/or the jet. Disentangling which site
dominates will be a major analysis task for the future.

Finally, we note that the presence of soft X-ray lines have been
subject to some doubt (e.g. \citealt{rutledge}) due to the relatively
poor statistics. Although the GRBs are relatively faint, the total
number of photons in the XMM-Newton detected X-ray lines are much
higher than in the claimed BeppoSAX features. For example, the claimed
iron line in GRB000214 (\citealt{antonelli}) has $\sim 35$ photons
whereas in GRB011211 there are $\sim 115$ line photons and $\sim 210$
in GRB030227. It would therefore seem perverse to accept BeppoSAX
detections but not those from XMM-Newton. More importantly, detailed
Monte-Carlo simulations (\citealt{reeves2}; \citealt{watson3}) give a
high probability that the observed line features are real. The null
probability for GRB011211 is about 4$\sigma$ whereas for GRB030227 it
is 4.4$\sigma$.

\subsection{GRB001025A} 

GRB001025A was observed by XMM-Newton some 45 hours after the burst.
The XMM-Newton data have poor S/N (Figure 5) but are best fitted by
a thermal plasma model providing a redshift of $z=0.53\pm0.03$
(\citealt{watson2a}). No optical redshift exists for this burst.

\subsection{GRB010220}

GRB010220 was observed by XMM-Newton some 15 hours after the burst. A
single powerlaw, with Galactic absorption, provides an acceptable fit
but adding a Gaussian emission line at $\sim 3.9$ keV improves the fit
at the 99\% level (Figure 6; \citealt{watson2a}). If identified with
the iron 6.4 keV K$\alpha$ line, this features gives an X-ray redshift
of $z=1.0\pm0.05$. Again no optical redshift has been determined for
this burst.

   \begin{figure}
   \centering
   \includegraphics[width=6.5cm]{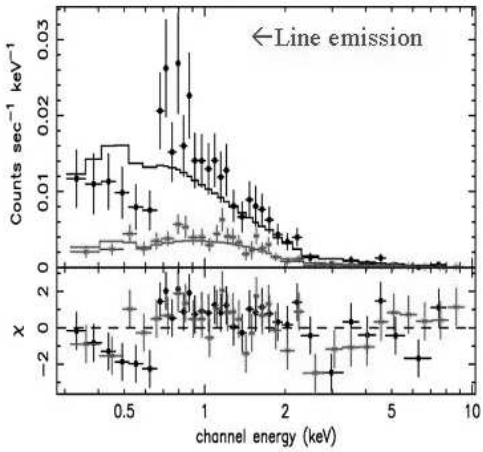}
      \caption{XMM-Newton EPIC PN and MOS spectra of GRB01025A fitted with a
powerlaw and Galactic absorption. There is an excess emission feature
observed at $\sim 0.8$ keV that can be well fitted with a thermal
plasma model (\citealt{watson2a}).
              }
\label{fig5}
   \end{figure}

   \begin{figure}
   \centering
   \includegraphics[width=6.5cm]{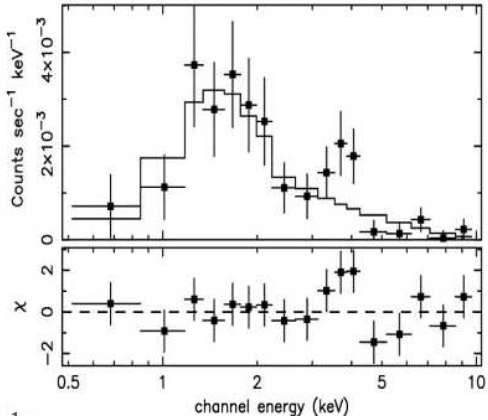}
      \caption{XMM-Newton EPIC PN spectrum of GRB010220 fitted with a
powerlaw and Galactic absorption. The only spectral feature is a possible
emission line, which could be due to iron K$\alpha$, at $\sim 3.9$ keV
(\citealt{watson2a}).
              }
\label{fig6}
   \end{figure}

\subsection{GRB020322}

This burst, discovered by BeppoSAX, shows soft X-ray absorption in
excess of that due to the Galaxy (Figure 7; \citealt{watson2b}).
Again, no optical redshift has been published, but the best-fit X-ray
absorption model, adopting either a neutral or ionised absorber, gives
a redshift of $z=1.8\pm1.0$. The best-fit absorbing column at $z=1.8$
is $1.3\pm0.2 \times 10^{22}$ cm$^{-2}$. There is no evidence for line
emission in these relatively high S/N data.

   \begin{figure}
   \centering
   \includegraphics[width=6.5cm]{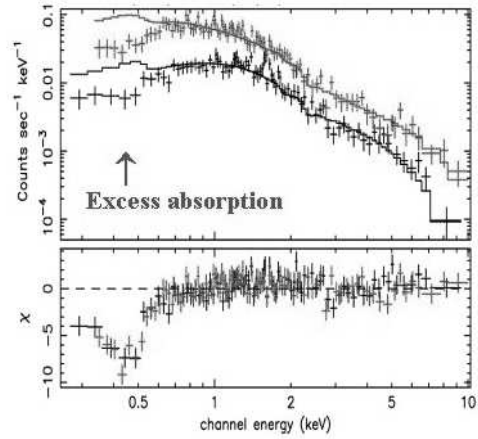}
      \caption{XMM-Newton EPIC PN and MOS spectra of GRB020322 fitted
with a powerlaw and Galactic absorption. There is an excess absorption
feature which can be well fitted with an absorber of redshift
$z=1.8\pm1.0$ (\citealt{watson2b}).
              }
\label{fig7}
   \end{figure}

\subsection{GRB030329}

The nearby ($z=0.1685$) GRB030329 is becoming a ``Rosseta Stone''
object. Its high initial apparent brightness and sky location allowed
for intensive, multi-waveband observations. Although unable to observe
it initially, XMM-Newton has provided invaluable late-time X-ray
fluxes (\citealt{tiengo}) which, when combined with other
multi-wavelength data, enabled \cite{willingale} to fit a
fireball shock model to parameterise the temporal behaviour with
unprecedented accuracy. The quality of the XMM-Newton spectra provide
an accurate measure of both the flux and the shape of the decaying
afterglow spectrum 37 and 61 days after the GRB. These spectra provide
an important constraint in the modelling. 

In Figure 8 we show the evolution of the X-ray and optical spectral
indices. For $t<1$ day the optical and X-ray indices differ by $\sim
0.5$ demonstrating that the coolling break occurs at $\sim
1.3\times10^{16}$ Hz. After a day the X-ray spectral index remains
constant while the optical spectrum reddens and is strongly reddened
for $t>10$ days. These changes are coincident with a rapid increase in
the optical flux at $t\sim1$ days relative to the X-ray (Figure 9)
which is only partly explained by the concurrent `hypernova'
(SN2003dh) seen in the optical (Hjorth et al. 2003). An``excess'' flux
is also seen in the radio (\citealt{willingale}). The excess energy
implies a longer-duration injection of energy than the initial GRB,
possibly due to a longer-lived central engine, an additional broader,
slower jet component and/or a supernova behaviour unlike others.
Whatever the explanation, XMM-Newton is uniquely powerful in allowing
a determination of both the X-ray flux and spectral shape at late
times.

   \begin{figure}
   \centering
   \includegraphics[angle=-90,width=6.5cm]{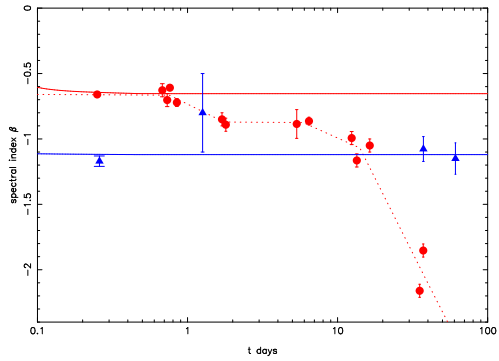}
      \caption{The observed X-ray (triangles) and optical (circles) spectral
indices ($f_{\nu} \propto \nu^{\beta}$) 
as a function of time for GRB030329 (\citealt{willingale} and
references therein). The dotted line indicates the trend in the
optical spectral index. The solid lines show the predicted evolution for 
the best-fit fireball shock model. }
\label{fig8}                            
   \end{figure}

   \begin{figure}
   \centering
   \includegraphics[width=6.5cm]{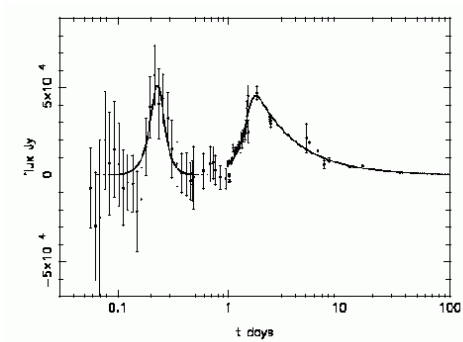}
      \caption{The 
residual optical (R band) flux after subtracting the best-fit fireball
shock model (\citealt{willingale}). The curves are simple powerlaw
fits for illustration.
}
\label{fig9}                            
   \end{figure}

\section{Conclusions}

From their discovery in the 1960s, the study of GRBs progressed
steadily until the discovery of afterglows. These fixed the distance
scale and hence the intrinsic luminosity of the bursts. The
observations of X-ray afterglows have revealed a wealth of spectral
detail only just beginning to be understood. XMM-Newton has already
made a major contribution to our understanding of GRBs, providing data
that challenge and in some cases contradict so-called ``standard
models''.

Although we have few objects to date, some interesting, possibly
related, spectral features are already starting to emerge. For
example, the first ``optical excess'' in GRB030329 occurs at $\sim
0.2$ days (rest-frame), intriguingly close to the rest-frame interval
at which time-dependent, soft X-ray emission-lines are seen in
GRB011211 and GRB030227. These events may indicate structure in the
jet and/or the surroundings close to the GRB. Obtaining
densely-sampled, correlated, multi-wavelength observations of such
events is obviously a vital task for the future. This task will be
made easier by the launch in 2004 of the dedicated GRB mission, Swift.
This mission will discover 2--3 GRBs per week, greatly increasing the
number of potential GRB targets for XMM-Newton. We hope XMM-Newton
will continue to observe as many GRBs as possible and hence further
our understanding.

\begin{acknowledgements}
We are grateful for the efforts of the XMM-Newton SOC in carrying out
the target of opportunity observations of GRBs.
\end{acknowledgements}

\bibliographystyle{aa}

\end{document}